\documentclass[useAMS,usenatbib]{mn2e}

\usepackage{natbib}
\usepackage{epsfig}
\usepackage{graphicx}

\title[Filamentary structures at $z \sim 0.8$]{Tracing the Filamentary 
Structure of the Galaxy Distribution at $z \sim 0.8$}

\author[Choi et al.]{Ena Choi$^{1}$\thanks{E-mail:echoi@astro.princeton.edu}, 
             Nicholas A. Bond$^{2}$,  
              Michael A. Strauss$^{1}$\thanks{E-mail:strauss@astro.princeton.edu}, 
              Alison L. Coil$^{3}$,
              \newauthor
              Marc Davis$^{4}$, 
              and 
              Christopher N. A. Willmer$^{5}$\\
$^1$Department of Astrophysical Sciences, Princeton University, 
     Princeton, NJ 08544, USA\\
$^2$Physics and Astronomy Department, Rutgers University, 
     Piscataway,  NJ 08854-8019, USA\\
$^3$Center for Astrophysics and Space Sciences (CASS), 
     Department of Physics, University of California, San Diego, 
     CA 92093, USA \\		
$^4$Department of Astronomy, University of California, Berkeley, 
         CA 94720, USA\\
$^5$Steward Observatory, University of Arizona, Tucson,
        AZ 85721, USA}

\begin{document}

\pagerange{\pageref{firstpage}--\pageref{lastpage}} \pubyear{2010}

\maketitle

\label{firstpage}

\newcommand{\beq}{\begin{equation}}
\newcommand{\eeq}{\end{equation}}
\def\mpch {$h^{-1}$ Mpc }
\def\magr {$M_{^{0.1}r}$ }
\def\boxsize {$320 \times 40 \times 14$ $\rm (h^{-1} Mpc)^3$ }

\begin{abstract}
We study  filamentary structure in the galaxy distribution at $z \sim 0.8$ 
using data from the Deep Extragalactic Evolutionary Probe 2 (DEEP2) 
Redshift Survey and its evolution to $z \sim 0.1$ using data from the 
Sloan Digital Sky Survey (SDSS). We trace individual filaments for both 
surveys using the Smoothed Hessian Major Axis Filament Finder, an 
algorithm which employs the Hessian matrix of the galaxy density field
to trace the filamentary structures in the distribution of galaxies. We 
extract 33 subsamples from the SDSS data with a geometry similar to 
that of DEEP2. We find that the filament length distribution has not 
significantly changed since $z \sim 0.8$, as predicted in a previous 
study using a $\Lambda$CDM cosmological N-body simulation. 
However, the filament width distribution, which is sensitive to the 
non-linear growth of structure, broadens and shifts to smaller widths
for smoothing length scales of $5-10$ \mpch from $z \sim 0.8$ to 
$z \sim 0.1$, in accord with N-body simulations.
\end{abstract}

\section{Introduction}\label{Intro}
The observed large-scale distribution of galaxies shows dense linear 
features:  filaments of galaxies which surround huge voids that appear 
largely empty, while rich clusters are found at their intersection 
\citep{dav82,del86,bon96,got05}. These structures are widely believed 
to have evolved through gravitational instability from small density 
fluctuations in the early universe. The evolution of large scale structure 
with cosmic time can probe the complex physics that governs the 
creation of galaxies in their host dark matter potential wells. A number of 
studies \citep[e.g.][]{coh96,con96,gia98,bro03,phl03,coi04b,ouc04,lef05,
men06} have  focused on the redshift dependence of the galaxy two 
point correlation function as a critical test of both cosmological and galaxy
evolution models. The comoving correlation length of galaxies is 
observed to be almost constant with redshift, which is interpreted as a 
consequence of the increasing bias of galaxies with redshift.

The two-point correlation function is a complete statistical measure of 
galaxy clustering only in the linear regime. Statistics of galaxy filaments, 
such as their lengths and widths, can be used as another useful tool to 
measure the large scale structure and test both cosmology and galaxy 
formation models.  Filaments,  with a typical length of $50-70$ \mpch 
\citep{bha04}, have been seen in every wide-field redshift survey, from 
the Great Wall of the CFA2 \citep{del86,gel89} to the very long filaments 
found \citep{got05} in the Sloan Digital Sky Survey (SDSS)  \citep{yor00}. 
Features qualitatively similar to the observed filamentary structures are 
also seen in numerical cosmological simulations. Various techniques 
have been proposed to identify and characterize filaments in 
observational and simulated samples \citep[e.g.][]
{moo83,eri04,sto05,lac05,nov06,ara07,sou08a,sou08b,sou09,for09}.
\citet[][hereafter Paper I and II, repectively]{bon09a,bon09b} use the 
eigenvectors of the Hessian matrix of the smoothed galaxy density field 
to identify and quantify filamentary structures. The filament length and 
width distributions of the observed local galaxy distribution from SDSS 
are consistent with those from N-body simulations at $z \sim 0$ 
adopting a standard cosmology. The time evolution of the filament 
network was studied in Paper II using cosmological N-body simulations; 
they found that the backbone of the filamentary structure is in place at 
$z=3$. These simulations show that non-linear growth of structure has 
little impact on the length of filamentary structures, but a great deal on 
the width. The dark matter filament width distribution evolves from 
$z\sim3$ to $z\sim0$, broadening and peaking at smaller widths as the 
universe expands.

Although many recent papers have studied the filamentary structures of 
local galaxy surveys \citep{sto07,sou08a,sou08b,gay09}, no study has 
been done on the evolution of filaments in redshift survey data. This is 
mainly due to the small volumes and the resulting severe cosmic 
variance of existing high redshift surveys. However, thanks to the 
successful completion of the Deep Extragalactic Evolutionary Probe 2
(DEEP2) Galaxy Redshift Survey \citep{dav03}, we can study the galaxy 
distribution at $z \sim 1$ over a large comoving volume ($5 \times 10^6$ 
$h^{-3} \rm Mpc^3$) over four widely separated fields.

In this study, we present measurements of filament statistics both for the 
galaxy distribution at $z \sim 0.8$ using the DEEP2 Galaxy Redshift 
Survey \citep{dav03}, which is an R-band-selected survey with a 
sampling density comparable to local surveys, and for the local galaxy 
distribution from the SDSS. We identify filaments in the galaxy 
distribution in DEEP2 using the Hessian matrix method of Paper I and II,
and draw subsamples from the SDSS redshift survey with the DEEP2 
geometry and sampling, to make a direct comparison between the two.

This paper is organized as follows:  \S~\ref{Method} summarizes the 
methods we use to find filamentary structures and measure their 
properties, referring the reader to Paper I and Paper II for more details.
In \S~\ref{sec:data}, we provide details of the data samples used here.
\S~\ref{Result} presents our results and \S~\ref{sec:summary} discusses 
their meaning and implications. We assume a flat concordance 
$\Lambda$CDM cosmology with $\Omega_m =0.3$, 
$\Omega_{\Lambda} = 1-\Omega_m = 0.7$ and $H_0 = 100$ $h$ km
s$^{-1}$ Mpc$^{-1}$ throughout this paper.

\section{Methods}\label{Method}
In this paper, we use the Smoothed Hessian Major Axis Filament Finder 
(SHMAFF), an algorithm that uses the eigenvectors of the Hessian matrix 
of the smoothed galaxy distribution to identify filamentary structures of 
galaxy data. The detailed methodology of SHMAFF and its applications 
are described in Papers I and II; Paper I quantifies the prominence and 
shapes of structures in the galaxy distribution using the Hessian matrix, 
while Paper II describes a method to find individual filaments, and 
compares their properties in cosmological N-body simulations to those 
in the SDSS galaxy distribution. We summarize the basics here. Since 
the geometry of the DEEP2 fields allows us to study filamentary 
structures best in two dimensions, we use a two-dimensional version of 
SHMAFF in this study.

\subsection{Smoothed density field and its Hessian}
To trace individual filaments in the galaxy distribution, we generate the 
density field and its second derivatives; filaments will be defined as 
regions with one eigenvalue of the Hessian matrix much larger than the 
other two. The density field is smoothed with a Gaussian kernel with 
smoothing length $l$, 
\begin{equation} 
\label{eq:Density} 
\tilde{\rho}({\bf x})=\int f({\bf x}-{\bf x'})\rho({\bf x'}){\rm d}^2{\bf x'}, 
\end{equation}
where $\rho(x)$ is the unsmoothed density field, and $f(x)$ is the 
smoothing kernel. In this study we use Gaussian smoothing,
\begin{equation}
f({\bf x})=\frac{1}{\sqrt{2\pi l^2}} e^{\frac{-|{\bf x}|^2}{2 l^2}},
\end{equation}
where $l$ is the smoothing length and the smoothing is performed over 
a two-dimensional box. The unsmoothed density field is given as a sum 
of delta functions at the positions of the galaxies and the smoothed 
Hessian (the matrix of second partial derivatives) is given by
\begin{equation} 
\label{eq:SmoothHessian} 
\tilde{H}_{{\rm ij}}({\bf x})=\int f({\bf x}-{\bf x'})\frac{{\rm \partial}^2 
\rho({\bf x'})}{{\rm \partial} x_{\rm i}'{\rm \partial} x_{\rm j}'}{\rm d}^2{\bf x'}
=-\int\frac{{\rm \partial}^2 f({\bf x}-{\bf x'})}{{\rm \partial} x_{\rm i}'{\rm \partial} 
x_{\rm j}'}\rho({\bf {\bf x'}}){\rm d}^2{\bf x'}.
\end{equation}

Since the filamentary structures can appear on a variety of scales, it is 
important to smooth on a series of length scales $l$ when searching for 
filaments.

The Hessian matrix describes the local curvature of the density field, that 
is, the major axis is aligned along the direction of lowest concavity. We 
compute the eigenvalues, $\lambda_i$ of the Hessian matrix  defined 
such that $\lambda_1<\lambda_2$, and eigenvectors, $\bf A_i$,  which 
give the orientation of the structure at a given grid cell. The direction of 
lowest concavity is expected to be along the filament itself. We thus 
simply need to find the major axis of the Hessian ellipsoid in order to 
find the direction of a filament at a given point in space. 

\subsection{Filament-Finding Parameters}
We trace filaments over those grid cells that satisfy the criteria 
$\lambda_1 < 0$, $\rho > \bar{\rho}$, where $ \bar{\rho}$ is the mean 
density of the smoothed density field. We choose a starting point at the 
local maximum density, and trace out the filament both parallel and 
antiparallel to the $\bf A_2$ axis until its local curvature exceeds a given 
threshold. Along a filament, if the angular rate of change of the axis of 
structure exceeds a value $C$, filament tracing is stopped and the point 
will be marked as a filament end. The stopping condition at pixel $m$ is 
given by
\begin{equation}
|{\bf A_{2,m}} \times {\bf A_{2,m-1}}| > {\rm sin}(C \Delta),
\label{eq:C}
\end{equation}
where $\Delta$ is the distance between pixels.

As each filament is found, the pixels associated with it are removed from 
further consideration as filament starting points. In particular, we 
introduce another input parameter $K$ and define a removal width 
$W_i$ at grid cell $i$ as
\begin{equation}
W_i = K \sqrt{\frac{-\rho_i}{\lambda_{1,i}}}.
\end{equation}
All pixels within a removal width $W_i$ of the most recently chosen 
filament element are excluded in the next iteration of the filament 
finding procedure, which prevents filaments from being multiply counted.
Here, a filament element is defined to be a segment of the filament with 
length equal to the distance between pixels.

This process depends on three input parameters: the smoothing length 
$l$ ($h^{-1}$ Mpc), the curvature criterion $C$ ($^{\circ} l^{-1}$) and the 
width of filament removal $K$. For tracing filaments in the large-scale
galaxy distribution, the best values of the input parameters suggested 
by Paper II are $C=30^{\circ} l^{-1}$ and  $40^{\circ} l^{-1}$ on 
smoothing scales of $l=5$ and 10 \mpch respectively, and $K=1$ for all
smoothing scales. These parameters were determined in three 
dimension; we adopt these values here in two dimension.

\subsection{Filament Measurement}
The length of each filament is defined as the distance along the filament 
between its two ends, which are specified by  Eq. \ref{eq:C} or where
the density no longer exceeds the threshold. As 
discussed in Paper II, the filament finder would identify an isolated 
spherical over-dense region as a "filament" of length of order the 
smoothing length. Thus in this paper, we exclude  "filaments" whose 
lengths are shorter than the smoothing length $l$.

The width of a filament element ($W$) is defined to be the root mean 
squared perpendicular offset of galaxies within a smoothing length $l$:
\begin{equation}
W = \sqrt{\frac{\Sigma_{i=1}^{N} |{\rm \bf R}_i|^2}{N}},
\end{equation}
where the sum is over the $N$ galaxies within a smoothing length of the 
filament element. Here, ${\rm \bf R}_i$ is the perpendicular offset of 
nearby points from the filament axis, and is defined as
\begin{equation}
{\rm \bf R}_i = {\rm \bf \hat{A}}_j \times 
       ({\rm \bf \hat{A}}_i \times ({ \rm \bf x}_j - { \rm \bf x}_i)),
\end{equation}
where ${\rm \bf \hat{A}}_j$ is the unit vector along the axis of structure.

\section{DATA}\label{sec:data}
The main data analyzed in this paper come from the DEEP2 and SDSS 
surveys. We now describe these two surveys to understand the samples 
and their selection funtions.

\subsection{DEEP2 Galaxy Redshift Survey}\label{sec:deep2}
The high-redshift galaxy sample used in this paper is from the DEEP2 
Galaxy Redshift Survey \citep{dav03}, which used the Deep Imaging and 
Multi-Object Spectrograph (DEIMOS) \citep{fab03} on the 10 m Keck II 
telescope to obtain spectra of optically selected galaxies at $z\sim1$. 
The selection was done from deep BRI photometry drawn from images 
taken with the CFHT12k camera on the Canada-France-Hawaii 
Telescope (CFHT) \citep{coi04b}. The DEEP2 spectra have a resolution 
of $R \sim 5000$, and rms redshift errors, which are determined from 
repeated observations, are $\sim 30$ $\rm km~s^{-1}$. The survey has 
measured high-confidence redshifts for $\sim 28,100$ galaxies in the 
redshift range of $0.7 < z < 1.5$  down  to a limiting magnitude of 
$R_{AB}=24.1$\footnote{All magnitudes in DEEP2 data are in the AB 
system. For photometric details, see \citet{coi04b}.}. The survey spans a 
comoving volume of approximately $5 \times 10^6$$h^{-3} \rm Mpc^3$, 
covering  $\sim3 ~\rm deg^2$ over four widely separated fields to limit 
the effect of cosmic variance. The DEEP2 observations, catalog 
construction, and data reduction are described in more detail in 
\citet{dav03,coi04a} and \citet{dav04}.

In fields 2, 3, and 4, the spectroscopic target galaxies are preselected 
using a color cut in $(B-R)$-$(R-I)$ space to ensure that most galaxies 
have redshifts greater than 0.75. With this color cut, $\sim 90$\% of the 
targeted galaxies are at $z>0.75$, and only $\sim3$\% of the $z>0.75$ 
galaxies brighter than the magnitude limit are not selected \citep{dav03}.
A fourth field, the extended Groth Strip (EGS), does not have this redshift 
preselection. We use the absolute B-band magnitudes ($M_B$) and 
restframe $U-B$ colors that are derived in \citet{wil06}; we apply no 
corrections for luminosity evolution. We create volume-limited samples 
as a function of  $M_B$ in three fields covering $\sim 2.2 ~\rm deg^2$.
The DEEP2 sample is not complete, and has a complicated angular 
mask. Details of sample definitions are discussed in Section
~\ref{sec:sample}.

\subsection{SDSS}

\begin{figure}
\begin{center}
\includegraphics[width=\columnwidth]{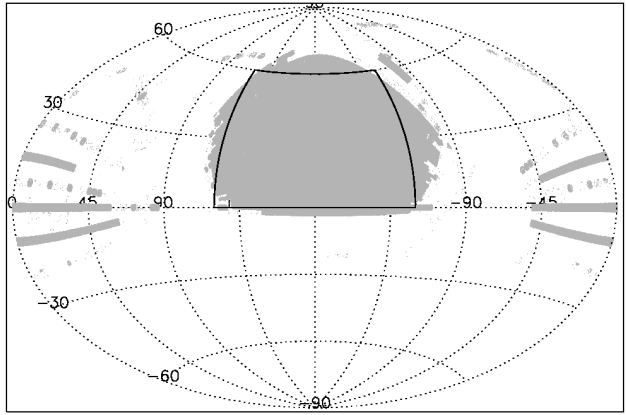}
\end{center}
\caption{ Aitoff projection in equatorial coordinates of the angular 
coverage of the SDSS Data Release 7 galaxy catalog \citep{aba09}.
Marked with a black solid line is the region from which samples used in 
this paper were drawn ($8^{\rm h} \le \alpha \le 16^{\rm h} $ and 
$0^{\circ} \le \delta \le 60^{\circ}$).}
\label{fig:Aitoff}
\end{figure}

We will compare the DEEP2 filamentary structures to those obtained from 
the SDSS at redshifts of $z \la 0.1$. The SDSS is an extensive 
photometric and spectroscopic survey, which has obtained photometry 
of a quarter of the sky and spectra of over 1.6 million objects. Imaging is 
obtained in the $u$, $g$, $r$, $i$, and $z$ bands \citep{fuk96, smi02, 
ive04} with a drift-scan camera with 30 2048$\times$2048 CCDs 
\citep{gun98} on a dedicated 2.5 m telescope \citep{gun06}. Spectra 
are measured with two fiber-fed digital spectrographs on the same 
telescope. Galaxies are selected for spectroscopy based on a 
magnitude limit \citep{str02}. An overview of the data pipelines and data 
products is provided in the Early Data Release  paper \citep{sto02}. 
The galaxy sample used in this paper are constructed from the SDSS 
Data Release 7 \citep[][hereafter DR7]{aba09}. As of DR7, the spectra of 
$\sim 930,000$ galaxies have been measured, covering 9380 
$\rm deg^2$. Galaxy redshift errors are typically $\sim 30 \rm km~s^{-1}$, 
which is similar to DEEP2.

For this analysis, we make use of the New York University Value Added 
Galaxy Catalog (NYU-VAGC), which is a compilation of the galaxy 
catalog from the SDSS DR7, publicly available at 
http://sdss.physics.nyu.edu/vagc/. A detailed description can be found in 
\citet{bla05}. We construct volume-limited samples from the northern 
portion ($8^{\rm h} \le \alpha \le 16^{\rm h} $ and $0^{\circ} \le \delta \le 
60^{\circ}$) of the survey as shown in Figure~\ref{fig:Aitoff}. We use 
$M_{^{0.1}r}$, the  $r$-band absolute magnitude corrected to its $z=0.1$ 
value using the K-correction code of \citet{bla03b} and the luminosity
evolution model of \citet{bla03c} to define volume-limited samples. We 
will define 33 subsamples from this SDSS sample volume shown in 
Figure~\ref{fig:Aitoff} to match the geometry of the DEEP2 samples.
Details of SDSS sample definitions and redshift range are discussed in 
Section~\ref{sec:sample}.

\subsection{Galaxy Sample Definitions}\label{sec:sample}
\subsubsection{Absolute magnitude cut}

\begin{figure}
\begin{center}
\includegraphics[width=\columnwidth]{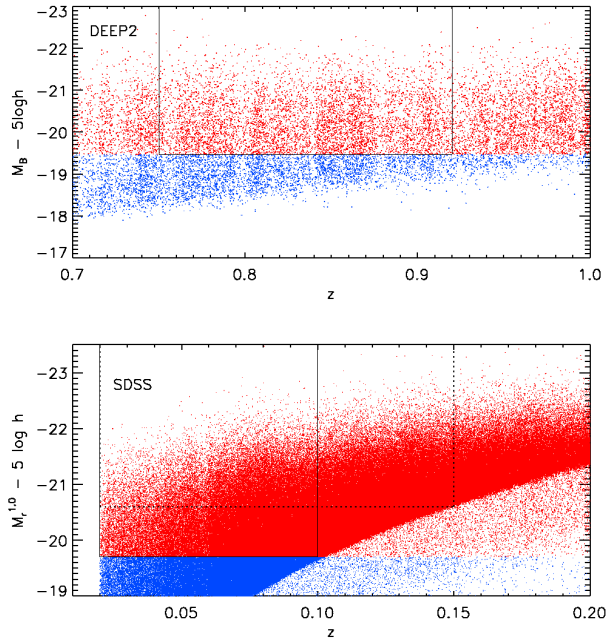}
\end{center}
\caption{
{\it (Top)} Absolute B-band magnitude (in AB magnitudes, with $h=1$) 
vs. redshift of the DEEP2 galaxy catalog. The magnitude and redshift
limits of the sample used here are shown in solid lines ($0.75\le z \le 
0.92$ and $M_B \le -19.47$). {\it (Bottom)} Absolute \magr magnitude 
(with $h=1$) of SDSS DR7 galaxy catalog as a function of redshift. The 
magnitude and redshift limits of the main sample used here are shown as 
solid lines ($0.02\le z \le 0.10$ and \magr $ \le -19.7$). The limits for the 
sparser sample used to study the effect of the direction of sample 
extraction (\S \ref{sec:subsample}) are shown as dotted lines ($0.02\le 
z \le 0.15$ and \magr $ \le -20.6$).
\label{fig:M_z}}
\end{figure}

As explained in Paper II, sparse sampling of the galaxy density field can 
impact the filament detection rate. A small number of galaxies per 
smoothing volume can create false filament detections and cause real 
filaments to be missed. In order to make the most direct comparison of 
the DEEP2 and SDSS galaxy distribution, we need to make sure they 
have the same densities.

The mean galaxy comoving number density of galaxies with absolute 
magnitudes below an absolute magnitude $M_{cut}$ is:
\begin{equation}
n_{\Phi} = \int_{-\infty}^{M_{cut}} \Phi(M) dM,
\label{eq:luminosity}
\end{equation}
where $\Phi(M)$ is the luminosity function of the galaxy survey in 
question. We adopt \citet{sch76} fits  from \citet{bla03a} for the SDSS 
galaxy luminosity function and from \citet{fab07} for the DEEP2 
luminosity function. We adjust the magnitude cut of each survey to define 
galaxy samples with matched number densities. We adopt the 
K-corrected rest-frame absolute magnitude cut $M_B = -19.47$ for 
DEEP2 and \magr=$-19.7$ for SDSS with $h=1.0$, and for which the 
two samples have comparable mean number densities of $n_{\Phi} 
\sim 0.008$ $h^3 \rm Mpc^{-3}$. These magnitude cuts are $\sim 1$ 
magnitude fainter than the characteristic magnitude $M^*$ of each 
survey. In Figure~\ref{fig:M_z}, we show the B-band absolute 
magnitude $M_B$ and redshift of each galaxy in the DEEP2 catalog 
and the regions from which our samples were drawn. The DEEP2 
sample is volume-limited for blue galaxies, but not for red galaxies, 
due to the selection in the observed R band, which corresponds to 
the rest-frame UV (see \citet{wil06} for more details on selection 
effects in the sample). Due to this  selection effect and the lower 
sampling density beyond $z\sim1$, we limit  our filamentary study 
in this paper to $z<0.92$.

Figure~\ref{fig:M_z} shows the absolute magnitude \magr of each 
galaxy from the SDSS catalog and cuts in magnitude and redshift with 
solid lines. Our SDSS sample consists of 528343 main sample galaxies 
\citep{str02} with $0.02 < z < 0.1$ and \magr$<-19.7$.

The magnitude cuts were chosen to make the two samples have 
comparable number density $n_{\Phi}$ based on the luminosity 
functions. However, the DEEP2 redshift sample is roughly $35$\% 
incomplete to its magnitude limit and redshift limit $0.75<z<0.92$, due to 
unobserved galaxies and redshift failures \citep{wil06} (this is in addition 
to the 3 \% incompleteness of the color selection, as described in
\S~\ref{sec:deep2}). We thus apply the DEEP2 
angular completeness window function to the SDSS sample to have the 
same number density ($n=0.005$ $h^3 \rm Mpc^{-3}$) as the observed 
DEEP2 sample. We do this by selecting subsamples of SDSS with the 
same geometry as DEEP2, as we describe in Section \ref{completeness} 
after we describe the DEEP2 geometry.

\subsubsection{Subsample definition}
\label{sec:subsample}
\begin{figure}
\begin{center}
\includegraphics[width=\columnwidth]{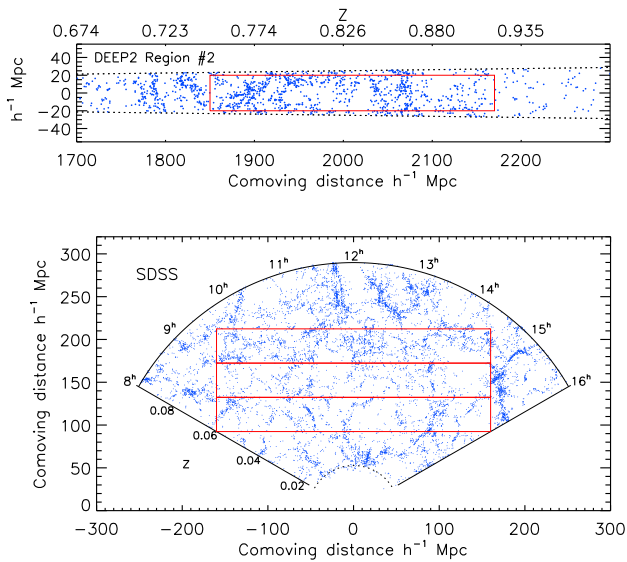}
\end{center}
\caption{ Example of the subsample extraction from the DEEP2 $\it (top)$
and SDSS  $\it (bottom)$ data. $\it Top:$ Redshift-space distribution of 
galaxies in Region 2 of DEEP2, shown as a function of redshift and 
comoving distance along the projected distance across the line of sight, 
assuming a $\Lambda$CDM cosmology. We extract one subsample from 
each region with dimensions of \boxsize (in red).
$\it Bottom:$ One slice with thickness of 14 \mpch from SDSS: extracted
subsamples are shown in red. We draw 33 such subsamples from the 
SDSS sample volume.
\label{fig:Extract}}
\end{figure}

Each DEEP2 field is much longer in the redshift direction than on the sky; 
the  $1-2\times0.5$ deg$^2$ fields used for this work span
$40-80\times20$ $h^{-1}$ Mpc  in transverse comoving extent, while the 
range $0.7<z<1.0$ corresponds to $560$ $h^{-1}$ Mpc  comoving in
the redshift direction.  From this comoving volume, we select subsamples 
with dimension of $\rm L_1 \times L_2 \times L_3$ = \boxsize=$179,200$ 
$h^{-3} \rm Mpc^3$, and define this as a standard box size. The comoving 
distance of the samples along the line of sight spans 1850 \-- 2170 
$h^{-1}$ Mpc, which corresponds to $0.75 < z< 0.92$. We extract three 
subsamples, one from each region (see the example in the upper panel of 
Figure \ref{fig:Extract}). We exclude the fourth field, the extended Groth 
Strip (EGS), which has a narrower width than other fields.

We can extract a larger number of subsamples with the standard box size 
out of the SDSS sample volume. We divide the volume into slices of 
thickness 14 $h^{-1}$ Mpc, and place as many rectangles into each slice 
as possible. Using this method, we obtain 33 subsamples in the redshift 
range of $0.02 < z < 0.10$. The bottom panel of Figure~\ref{fig:Extract} 
shows an example of subsample extraction in an SDSS slice.

\begin{figure}
\begin{center}
\includegraphics[width=\columnwidth]{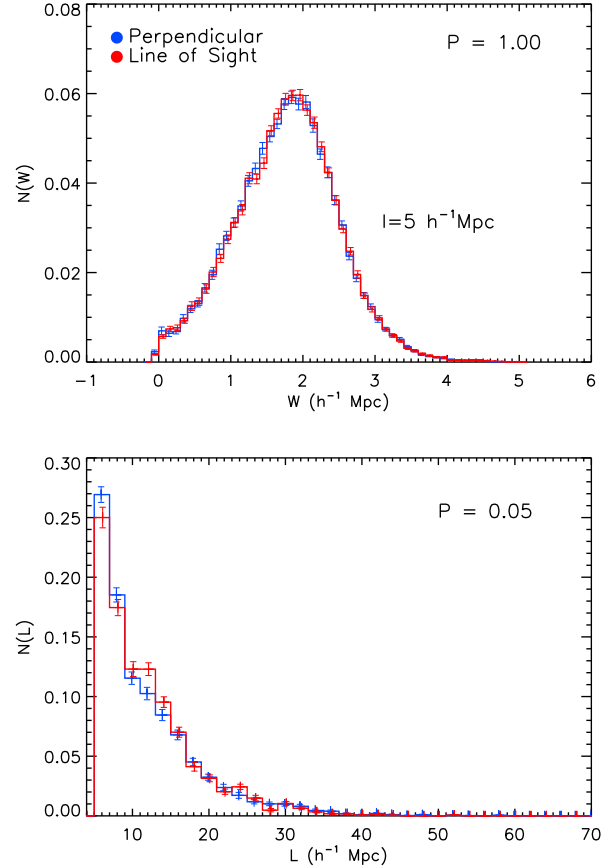}
\end{center}
\caption{
The width distributions {\it (top)} and the length distribution {\it (bottom)}
of filaments of subsamples perpendicular (in blue) and parallel (in red) 
to the line of sight found with the smoothing length $l =5$ $\rm h^{-1} 
Mpc$, $C=30^{\circ} l^{-1}$, and $K=1$.
\label{fig4}}
\end{figure}

The long axis of the DEEP2 subsamples is along the redshift direction, 
while it is perpendicular to it in SDSS. One might be concerned that 
redshift distortions due to peculiar velocities along the line of sight would
affect the filamentary statistics differently in the SDSS and DEEP2 
subsamples. We were forced to do this, as the volume-limited SDSS 
sample has an extent in the radial direction less than the length of the 
standard box. We test the effect of the direction of sample extraction
by comparing the filamentary properties from the subsamples extracted
perpendicular to the line of sight to those extracted parallel to the
line of sight. To do this, we build another sparser sample from the SDSS 
over the redshift range of $0.02 < z < 0.15$, with an absolute magnitude 
cut $M_{r^{0.1}} < -20.6$ (Figure~\ref{fig:M_z}). We extracted 135 
subsamples perpendicular to the line of sight  and 58 subsamples 
parallel to the line of sight. The filaments are found with the smoothing 
length $l =5$ $\rm h^{-1} Mpc$, $C=30^{\circ} l^{-1}$, and $K=1$. We 
derive the length and width distributions for each subsample and 
calculate the composite distributions and its error by calculating the mean 
and a standard deviation at each bin. As shown in Figure~\ref{fig4}, we 
found that the filament width distributions are essentially 
indistinguishable between the perpendicular and parallel subsamples.
This is confirmed by the Kolmogorov-Smirnov two-sample test (K-S test),
which we applied  to the two distributions. The length distributions of the
two are essentially identical, differing by only 2 $\sigma$ by the K-S test. The effect 
of the direction of the subsample extraction on the filament finder and 
filamentary properties are negligible presumably because we smooth the 
density field with a smoothing length that is much larger than typical 
galaxy peculiar velocities.

\begin{table*}
   \begin{center}
   \caption{Summary of filamentary property}
   \vskip+0.5truecm
   \begin{tabular}{c|c|c|c|c}\hline\hline
  Smoothing length     &  \multicolumn{2}{|c|}{5 \mpch} & 
  \multicolumn{2}{|c}{10 \mpch} \cr 
  \hline
  data & DEEP2 &    SDSS   & DEEP2 & SDSS \cr
  \hline\hline
  Total number of filaments & 97&958&24&245\cr
  mean $\#$ of filaments per subsample& $32.3 \pm 4.7$ & $29.0 \pm4.9$
  & $8.0\pm1.7$ & $7.42\pm2.2$ \cr
 mean total length of filaments ($h^{-1}$ Mpc)  & $404.0\pm29.5$  & 
 $376.1\pm53.8$ & $216.6\pm20.0$ &$186.5\pm43.4$ \cr
  \hline\hline
 width distribution peak $\mu$ ($h^{-1}$ Mpc) & $2.021 \pm 0.009$ & 
 $1.924 \pm 0.006$ & $3.963 \pm 0.023$ & $3.802 \pm 0.017$ \cr
 width distribution $\sigma$ ($h^{-1}$ Mpc) & $0.508 \pm 0.005$ & 
 $0.544 \pm 0.004$& $0.580 \pm 0.016$ & $0.808 \pm 0.011$ \cr
  \hline\hline
  \end{tabular}
  \end{center}
  \label{tab1}
\end{table*}

\subsubsection{Survey completeness}\label{completeness}
We have to consider the survey completeness.  The DEEP2 survey
spectroscopically targets $\sim$60\% of objects that pass the apparent
magnitude and color cuts mentioned above.  The redshift success is 
$73 \%$ of those targeted galaxies \citep{wil06}. Therefore, we have 
successful redshifts for $\sim50$\% of all galaxies in the surveyed fields 
with apparent magnitude of $R<24.1$.  The sampling rate is a complex 
function of position across each field. In order to model this effect, we 
use the angular window function of the DEEP2 survey. The mask gives 
the completeness at each angular position; in unobserved regions such 
as around bright stars, the completeness is zero. We project the angular 
window function covering the three DEEP2 regions onto the geometry of
each box to generate three-dimensional completeness maps. Then each 
SDSS subsample with the geometry of the DEEP2 standard box is diluted 
with a completeness map randomly selected among the three DEEP2 
samples.

We have further complications because the sampling rate is 
non-uniform due to the necessities of slitmask design. Spectra of 
objects are not allowed to overlap on the CCD when observing with 
multi-object slit masks, therefore, objects that lie near each other in the 
direction on the sky that maps to the wavelength direction on the CCD 
cannot be  observed simultaneously. This results in under-sampling in 
the highest density regions  on the plane of the sky. In order to reduce 
the effect of this bias, adjoining slit masks are positioned approximately 
a half-mask width apart, giving each galaxy at least two chances to be 
on a mask \citep{coi08}. Despite this, the probability that a target with 
nearest neighbor $<10"$ away is selected for spectroscopy is 
diminished by $\sim 25 \%$. Many DEEP2 related papers 
\citep{coi04a,con05} model this effect by applying the actual DEEP2 
mask-making algorithm to the mock galaxy catalogs, which throws out 
some galaxies located close to other galaxies in the sky. This effect on 
the filamentary properties is negligible, as it is more relevant on small 
scales $\la 2$  $\rm h^{-1} Mpc$ \citep{coi06}. In particular, \citet{coi08}
have studied the bias in the two-point correlation function due to the
slitmask effect; they found it was 3.5 \% at 1 $h^{-1}$ Mpc, 1 \% 
at 5 $h^{-1}$ Mpc (the minimum smoothing length we use), and 
under a percent by 10 $h^{-1}$ Mpc.

The final SDSS subsamples have a mean of 992 galaxies each, with a 
standard deviation of 220, comparable to the DEEP2 values (1259, 616, 
and 1095 galaxies in each field). The mean number density of galaxies 
in the DEEP2 and SDSS subsamples is $n \sim 0.0055$ $h^3 \rm 
Mpc^{-3}$ with a standard deviation of $\sim 0.001$ $h^3 \rm Mpc^{-3}$.

\section{Results}\label{Result}

\begin{figure*}
\begin{center}
\includegraphics[width=\textwidth]{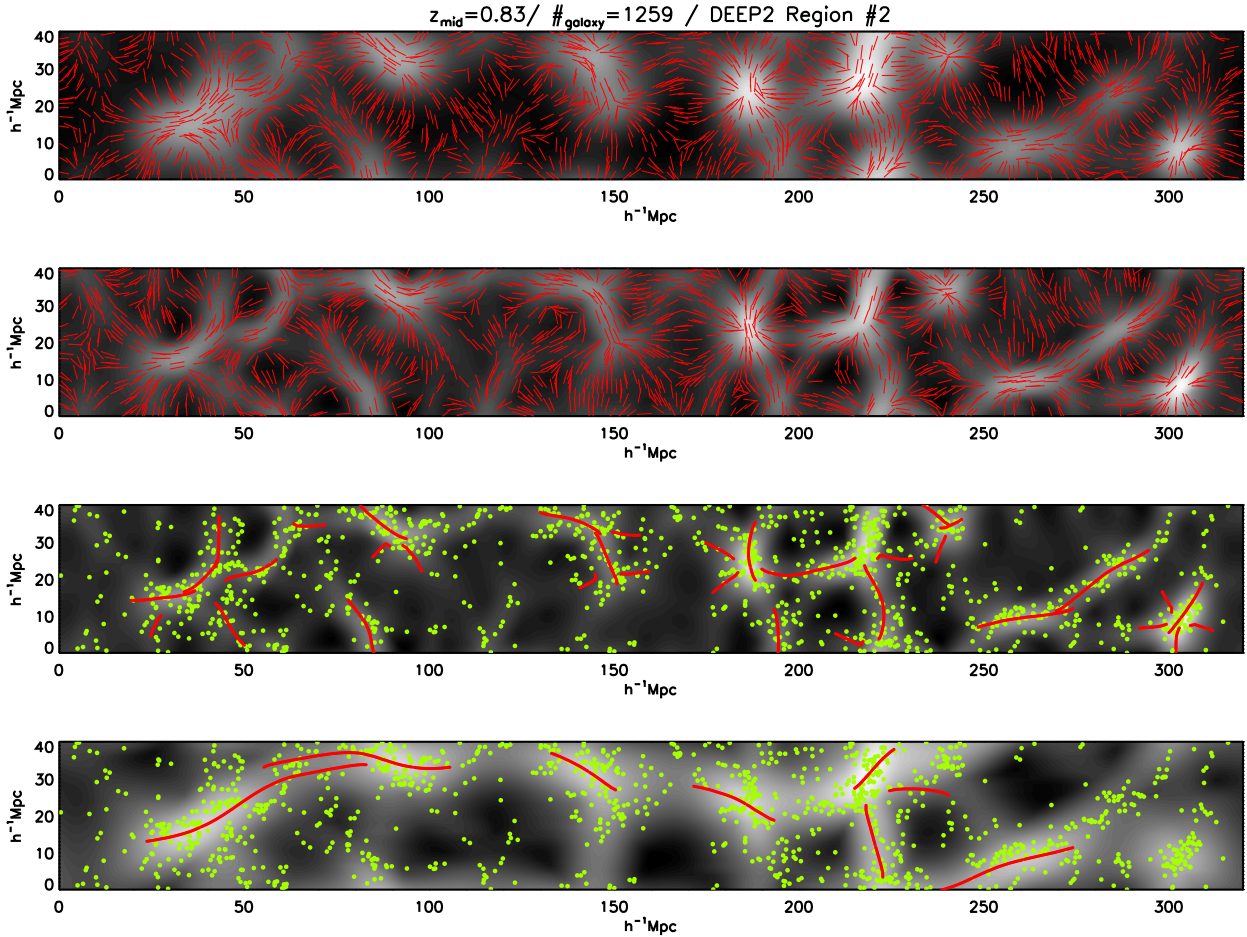}
\end{center}
\caption{The filament finder algorithm is shown in action for the 
subsample that corresponds to DEEP2 Region 2,  a \boxsize slice in the 
redshift range $0.75<z<0.92$.  The smoothing length is $l =5$ 
$\rm h^{-1} Mpc$. (a) An inverted grayscale density map (lighter=more 
dense) along with a 20\% random sample of bars that indicate the 
direction of the axis of structure at each point. (b) a grayscale map of 
$\lambda_1$ (lighter=more negative), the first eigenvalue of the second 
partial derivative of the density field. (c) The galaxy distribution (green 
dots) and filaments (in red) are shown  with the grayscale map of 
$\lambda_1$ (lighter=more negative).   The filaments are found with 
parameters $C=30^{\circ} l^{-1}$ and $K=1.0$. (d) Same as (c) but for the 
smoothing length $l =10$ $\rm h^{-1} Mpc$. The filaments are found with 
parameters $C=40^{\circ} l^{-1}$ and  $K=1.0$.
\label{fig:bar}}
\end{figure*}

We project the galaxy distribution in each of the three DEEP2 
subsamples and 33 SDSS subsamples along the short axis and 
measure a two-dimensional density field. We smooth each subsample 
with smoothing lengths $l=5$ and $10$ \mpch and run the filament finder 
on them using $C=30^{\circ} l^{-1}$ and $40^{\circ} l^{-1}$ respectively. 
The width of filament removal  is given as $K=1$ for all smoothing scales.
Figure~\ref{fig:bar}  shows the density field smoothed with $l=5$ $h^{-1}$ 
Mpc, maps of $\lambda_1$ and the identified filaments for DEEP2 Region 
2. In the top two panels, the red bars indicate the direction of the axis of 
structure at each point. The axis of structure aligns with the local 
filamentary structure, and we identify individual filaments using the 
curvature criterion $C=30^{\circ} l^{-1}$ and $K=1$. In the bottom two 
panels of Figure~\ref{fig:bar}, we show the identified filaments along with
the galaxy distribution itself in green dots for the two smoothing lengths.

After excluding filaments shorter than a smoothing length $l$, we found 
97 filaments in the three DEEP2 subsamples, and 958 filaments in the 33 
SDSS subsamples, with smoothing length of $l=5$$h^{-1}$ Mpc. The 
mean total length of filaments of each subsample is $404.0$ \mpch with a 
standard deviation of $29.5$ \mpch for DEEP2, and $376.1$ \mpch with a 
standard deviation of $53.8$ \mpch for SDSS. This gives a total filament 
length per unit area of $3.16 \times 10^{-2}$ $h$ $\rm Mpc^{-1}$ for 
DEEP2 subsamples, and $2.94 \times 10^{-2}$ $h$ $\rm Mpc^{-1}$ for 
SDSS. These results are in excellent agreement; the overall length of 
filaments in the two cases is indistinguishable. We summarize the 
filamentary properties of DEEP2 and SDSS at the two smoothing length 
scales  in Table 1.

\begin{figure*}
\begin{center}
\includegraphics[width=\textwidth]{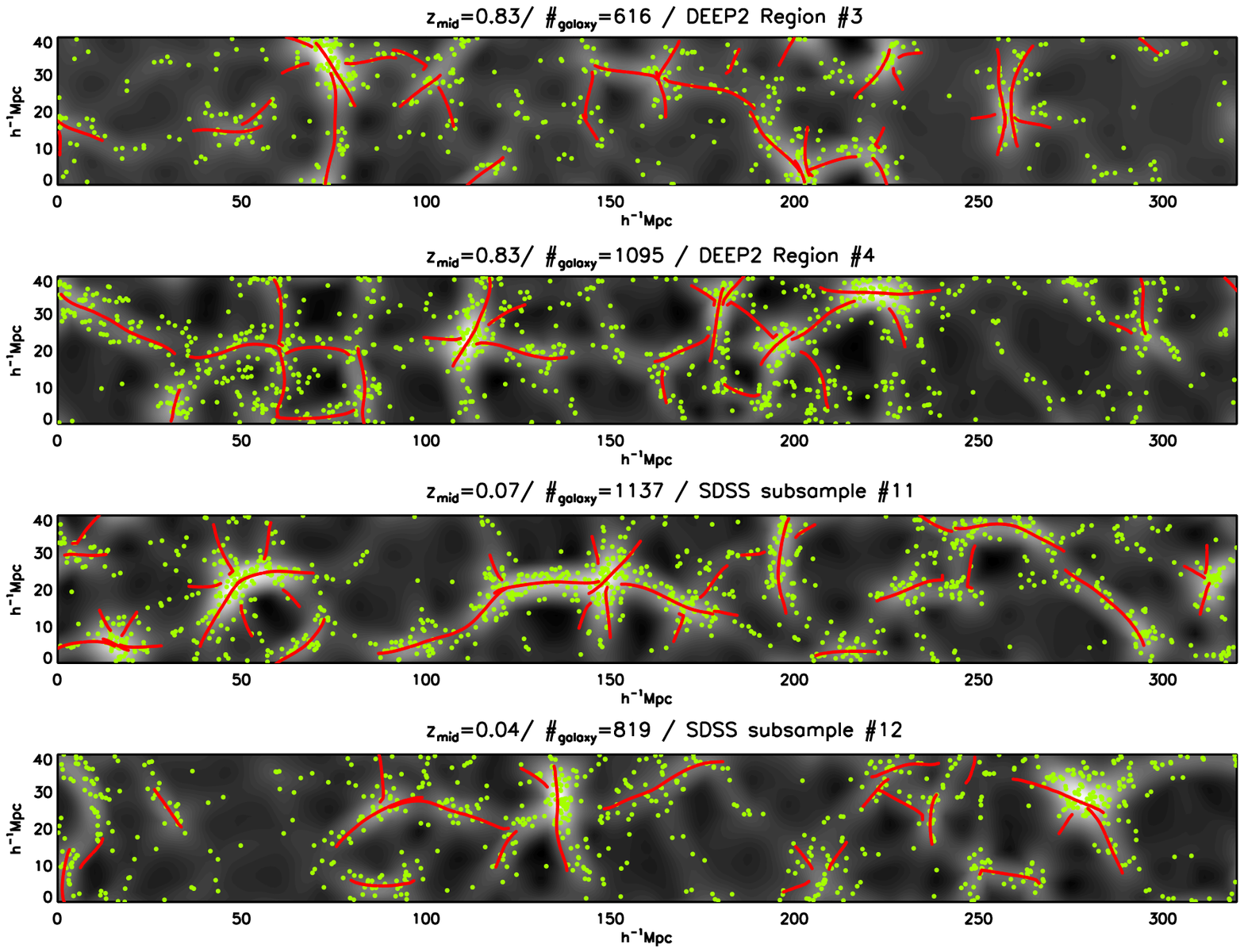}
\end{center}
\caption{The distribution of filaments in the DEEP2 (upper two panels)
and two randomly chosen SDSS subsamples (lower two panels). The 
smoothing length is $l =5$ $\rm h^{-1} Mpc$ and the  filaments are found 
with filament-finding parameters $C=30^{\circ} l^{-1}$ and $K=1.0$. The 
grayscale map of $\lambda_1$ (lighter=more negative) is shown along 
with the galaxy distribution (green dots) and filaments (in red).
\label{fig:filament}}
\end{figure*}

In Figure~\ref{fig:filament}, we show the filaments in Region 3 and 4 of 
DEEP2 and two selected subsamples of SDSS. The grayscale map of
$\lambda_1$ is shown with the identified filaments in red and the galaxy 
distribution in green dots. The filament distributions of DEEP2 and 
SDSS subsamples look qualitatively similar. We count how many 
galaxies lie within one smoothing length ($l$) of a filament,
and thus calculate the fraction of galaxies within a smoothing length of a 
filament for two surveys. With smoothing length $l =5$ $\rm h^{-1} Mpc$, 
$81.2 \pm 1.9$\% of DEEP2 galaxies and $82.1\pm0.5$\% for SDSS 
galaxies lie within one smoothing length of filaments; the fraction of 
filament galaxies is almost identical in the two surveys.

\begin{figure}
\begin{center}
\includegraphics[width=\columnwidth]{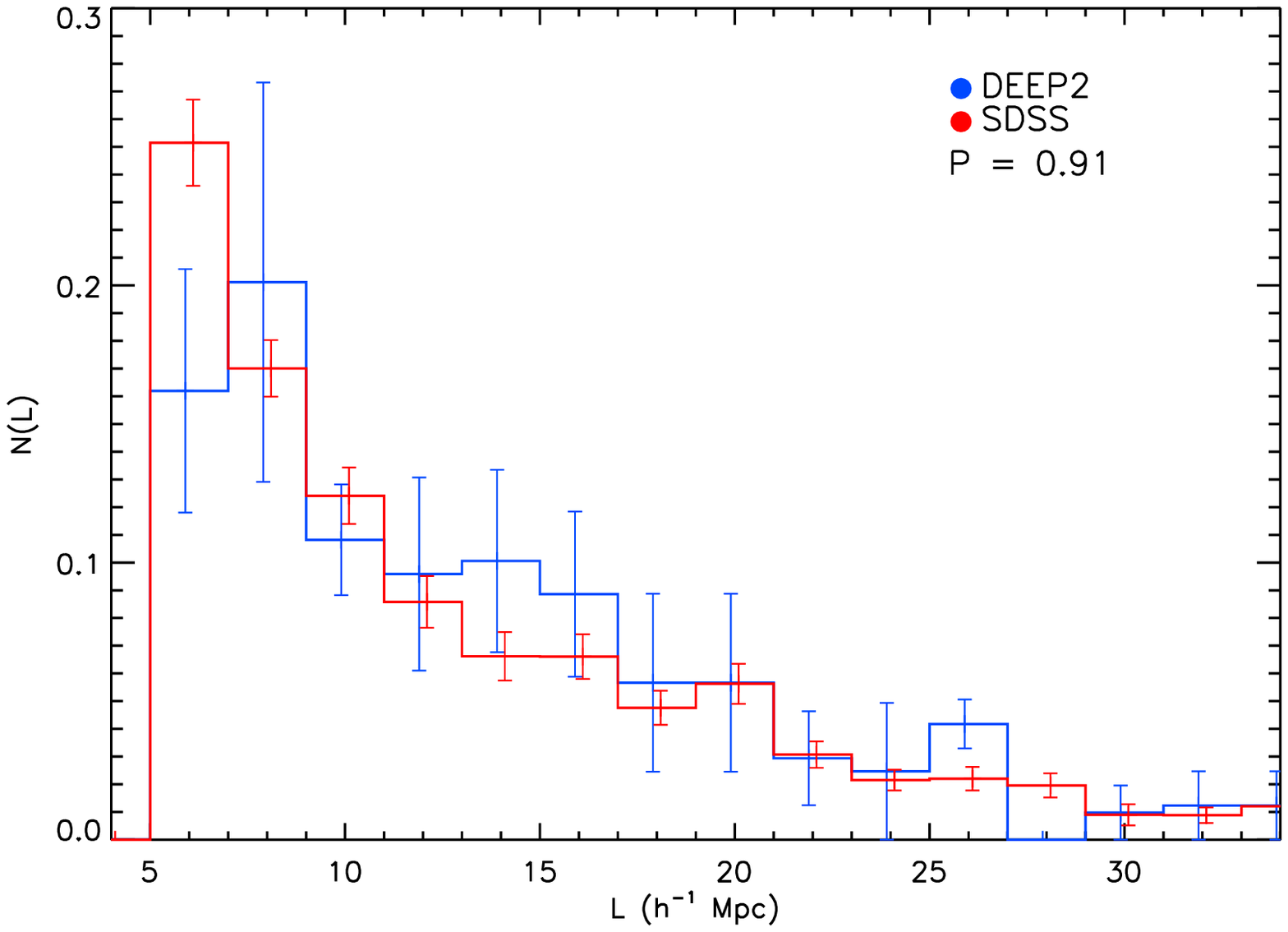}
\end{center}
\caption{The composite length distributions of the filaments of
three DEEP2 subsamples (in blue) and 33 SDSS subsamples
(in red).
\label{fig:length}}
\end{figure}

Figure~\ref{fig:length} shows the length distribution of filament for SDSS 
and DEEP2 with $l=5$ \mpch smoothing. There are too few filaments at 
10 \mpch smoothing to make a useful comparison. We derive the length
distribution for each subsample and calculate the composite distribution 
and its error by calculating the mean and a standard deviation at each 
bin. The length distributions found in DEEP2 are similar to those found in 
SDSS. The numerical simulations in \citet{bon08} showed that dark 
matter filaments have an exponential length distribution at large filament 
lengths that very closely matched that found in a Gaussian random field
with the same power spectrum. This means that even if the filaments in 
the two galaxy distributions are at different stages of their evolution, the 
length distribution should be similar between the two.

\begin{figure}
\begin{center}
\includegraphics[width=\columnwidth]{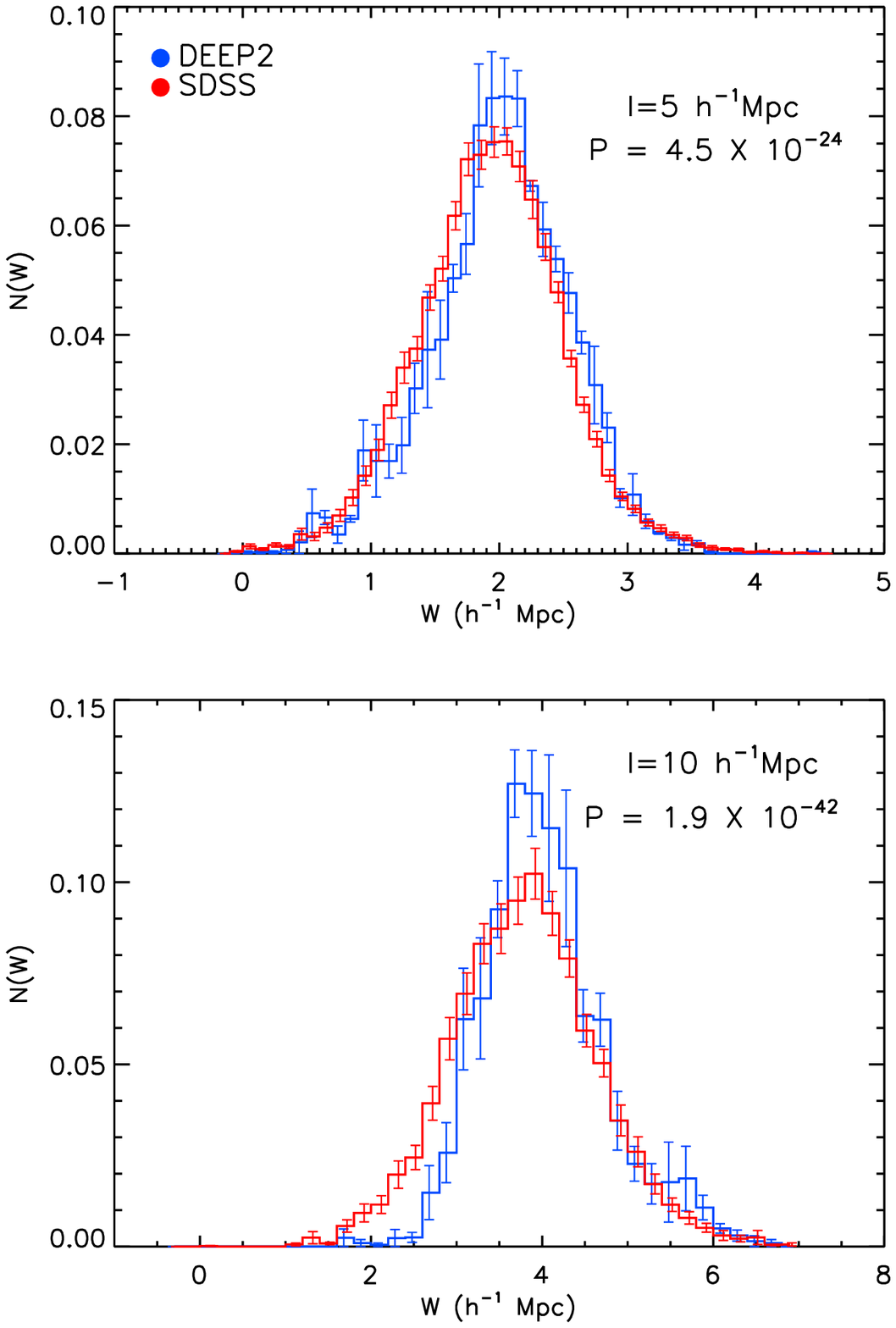}
\end{center}
\caption{The composite width distributions of the filaments of three 
DEEP2 subsamples (in blue) and 33 SDSS subsamples (in red).
\label{fig:width}}
\end{figure}

However, the width distribution of filament elements changes significantly 
as non-linear evolution proceeds (Paper II): it broadens and peaks at 
smaller widths  with cosmic time, according to N-body simulations. We 
show the width distributions of filament elements for DEEP2 and SDSS in 
Figure~\ref{fig:width} for the  two smoothing lengths. We applied a K-S 
two-sample test to compare the two; the calculated probability that the two 
are drawn from the same distribution is shown in each panel in 
Figure~\ref{fig:width}. In both cases, the probability is negligibly small. We 
make Gaussian fits to the distributions to compare the two surveys. 
For $l=5$ $h^{-1}$ Mpc, DEEP2 has a width distribution that peaks at 
$\mu=2.021 \pm 0.009$ \mpch with $\sigma=0.508 \pm 0.005$$h^{-1}$ 
Mpc, and SDSS has $\mu=1.924 \pm 0.006$ \mpch with $\sigma=0.544 
\pm 0.004$ $h^{-1}$ Mpc. In case of $l=10$ $h^{-1}$ Mpc, the distribution 
has $\mu=3.963\pm 0.0023$ \mpch with $\sigma=0.580 \pm 0.016$ 
\mpch for DEEP2, and $\mu=3.802 \pm 0.017$ \mpch and $\sigma=0.808 
\pm 0.011$ \mpch for SDSS. For both smoothing lengths, the filament 
element width distributions broaden and shift to smaller widths from 
$z\sim 0.8$ to $z \sim 0.1$. Figure \ref{fig:width} shows better match 
between the width distributions of SDSS and DEEP2 at high widths 
than low, in agreement with the simulations shown by \citet{bon08}. Thus, 
the widest filaments narrow more slowly than do the narrowest ones, 
as one would expect in the ellipsoidal collapse model \citep{zel70},
in which overdensities sequentially contract along their principal axes, 
in order of increasing length.

\section{Summary}\label{sec:summary}
We study the time evolution of the filament network in the galaxy 
distribution by comparing the filamentary structure at $z \sim 0.8$ from 
the Deep Extragalactic Evolutionary Probe 2 (DEEP2) Redshift Survey 
and those at $z \sim 0.1$  from the Sloan Digital Sky Survey (SDSS).
We trace individual filaments for both surveys using SHMAFF, an 
algorithm which employs the Hessian matrix to trace the filamentary 
structures in the distribution of structure. We define three subsamples 
from DEEP2 and 33 subsamples from SDSS, with the same sampling 
and geometry, namely a box of $320 \times 40 \times 14$ 
$\rm (h^{-1} Mpc)^3$. We smooth the galaxy distribution with length 
scales of $l=5$ and $10$ $h^{-1}$ Mpc, and trace individual filaments 
along the axis of structure, and mark the end of filaments when the axis 
orientation changes more rapidly than a preset threshold of $C=30$ and 
$40$$^{\circ} l^{-1}$ respectively. We found 97 filaments in DEEP2 
subsamples and 957 filaments in SDSS subsamples with smoothing 
length $l=5$ $h^{-1}$ Mpc, and 24 filaments for DEEP2 and 230 for 
SDSS for $l=10$$h^{-1}$ Mpc. Thus the number of filaments per unit
volume is unchanged from high to low redshift. We find that filament 
length distribution has not changed significantly since $z \sim 1$, 
however, the filament width distribution, which is sensitive to non-linear 
growth of structure, broadens and shifts to smaller widths for smoothing 
length scales of $5$ and $10$ \mpch from $z \sim 0.8$ to $z \sim 0.1$.
The evolution in the length and width distributions is consistent with 
predictions from a $\Lambda$CDM cosmological N-body simulation. 
As found in Paper II, non-linear growth of structure has a great impact on 
the width of filamentary structures.

We restricted our study to two-dimensional analysis due to the the 
geometry of the DEEP2 survey. In order to better show the filamentary 
evolution, however, the filamentary structures should be studied in larger 
volumes and analyzed in three dimensions. The next generation of galaxy 
surveys, which will target the early universe with larger volume and depth
can open up the possibility of detailed study of filamentary structures and 
their evolution. These surveys include the Advanced Dark Energy Physics 
Telescope (ADEPT), a space-based spectroscopic survey that promises to
determine the location of 100 million galaxies at $1 < z < 2$,  the BigBOSS 
\citep{sch09}, a proposed ground-based wide field spectroscopic survey
at $0.2<z<3.5$, and all hemisphere HI redshift surveys with the Square 
Kilometer Array (SKA) \citep{raw04}.

\section*{Acknowledgments}
We thank Jeffrey Newman, Michael Blanton, Charlie Conroy and the referee,
Michael Vogeley for useful discussions.

Funding for the creation and distribution of the SDSS Archive has been 
provided by the Alfred P. Sloan Foundation, the Participating Institutions, 
the National Aeronautics and Space Administration, the National Science 
Foundation, the U.S. Department of Energy, the Japanese 
Monbukagakusho and the Max Planck Society. The SDSS Web site is 
http://www.sdss.org/.

The SDSS is managed by the Astrophysical Research Consortium (ARC) 
for the Participating Institutions. The Participating Institutions are The 
University of Chicago, the Fermi National Accelerator Laboratory 
(Fermilab), The Institute for Advanced Study, The Japan Participation 
Group, The Johns Hopkins University, The Korean Scientist Group, the 
Los Alamos National Laboratory, The Max-Planck-Institute for Astronomy 
(MPIA), The Max-Planck-Institute for Astrophysics (MPA), the New Mexico 
State University, The University of Pittsburgh, Princeton University, The 
United States Naval Observatory, and The University of Washington. 
E. C. and M. A. S. acknowledge the support of NSF grant AST-0707266.


\begin{thebibliography}{}

\bibitem[Abazajian et al.(2009)]{aba09}
Abazajian, K. N. et al., 2009, ApJS, 182, 543

\bibitem[Arag{\'o}n-Calvo et al.(2007)]{ara07}
Arag{\'o}n-Calvo, M. A., Jones, B. J. T., van de Weygaert, R., \& 
van der Hulst, J. M. 2007, A\&A, 474, 315

\bibitem[Bharadwaj, Bhavsar \& Sheth(2004)]{bha04}
Bharadwaj, S., Bhavsar, S. P., \& Sheth, J. V. 2004, ApJ, 606, 25

\bibitem[Blanton et al.(2003a)]{bla03a}
Blanton M. R., et al. 2003a, ApJ, 592, 819

\bibitem[Blanton et al.(2003b)]{bla03b}
Blanton, M. R., et al. 2003b, AJ, 125, 2276

\bibitem[Blanton et al.(2003c)]{bla03c}
Blanton M. R., et al. 2003c, AJ, 125, 2348

\bibitem[Blanton et al.(2005)]{bla05}
Blanton M. R., et al. 2005, AJ, 129, 2562

\bibitem[Bond \& Myers(1996)]{bon96}
Bond, J. R., \& , Myers, S. T. 1996, ApJS, 103, 1

\bibitem[Bond(2008)]{bon08}
Bond, N. A. 2008, Ph.D. thesis, Princeton University
  
\bibitem[Bond, Strauss, \& Cen(2009)]{bon09a}
Bond, N. A., Strauss, M. A., \& Cen, R. 2009, submitted to
MNRAS (arXiv:0903.3601) (Paper I)
  
 \bibitem[Bond, Strauss, \& Cen(2010)]{bon09b}
Bond, N. A., Strauss, M. A., \& Cen, R. 2010, submitted to
MNRAS (Paper II)
 
\bibitem[Brown et al.(2003)]{bro03}
Brown, M. J. I., Dey, A., Jannuzi, B. T., Lauer, T. R., Tiede, G. P.,
  \& Mikles, V. J. 2003, ApJ, 597, 225

\bibitem[Cohen et al.(1996)]{coh96}
Cohen, J. G., Hogg, D. W., Pahre, M. A., \& Blandford, R. 1996, 
ApJ, 462, L9

\bibitem[Coil et al.(2004a)]{coi04a}
Coil, A. et al. 2004, ApJ, 609, 525

\bibitem[Coil et al.(2004b)]{coi04b}
Coil, A. et al. 2004, ApJ, 617, 765

\bibitem[Coil et al.(2006)]{coi06}
Coil, A. et al. 2006, ApJ, 638, 668
  
\bibitem[Coil et al.(2008)]{coi08}
Coil, A. et al. 2008, ApJ, 672, 153

\bibitem[Connolly et al.(1996)]{con96}
Connolly, A.et al. 1996, ApJ, 473, L67

\bibitem[Conroy et al.(2005)]{con05}
Conroy, C. et al. 2005, ApJ, 625, 990
  
\bibitem[Davis et al.(1982)]{dav82}
Davis, M., Huchra, J., Latham, D. W., \& Tonry, J. 1982, ApJ, 253, 423

\bibitem[Davis et al.(2003)]{dav03}
Davis, M. et al. 2003, Proc. SPIE, 4834, 161

\bibitem[Davis, Gerke, \& Newman(2004)]{dav04}
Davis, M., Gerke B. F., \& Newman, J. A. 2004, preprint 
(arXiv:astro-ph/0408344)

\bibitem[de Lapparent, Geller, \& Huchra(1986)]{del86}
de Lapparent, V., Geller, M. J., \& Huchra, J. P., 1986, ApJ, 302, L1

\bibitem[Eriksen et al.(2004)]{eri04}
Eriksen, H. K., Novikov, D. I., Lilje, P. B., Banday, A. J., \& G{\'o}rski, 
K. M. 2004, ApJ, 612, 64 

\bibitem[Faber et al.(2003)]{fab03}
Faber, S., et al. 2003, Proc. SPIE, 4841, 1657

\bibitem[Faber et al.(2007)]{fab07}
Faber, S., et al. 2007, ApJ, 665, 265

\bibitem[Forero-Romero et al.(2009)]{for09}
Forero-Romero, J. E., Hoffman, Y., Gottloeber, S., Klypin, A., \& 
Yepes, G. 2009, MNRAS, 396, 1815

\bibitem[Fukugita et al.(1996)]{fuk96}
Fukugita, M., Ichikawa, T., Gunn, J. E., Doi, M., Shimasaku, K., \& 
Schneider, D. P. 1996, AJ, 111, 1748

\bibitem[Gay et al.(2009)]{gay09}
Gay, C., Pichon, C., Le Borgne, D., Teyssier, R., Sousbie, T., \& 
Devriendt, J. 2009, arXiv:0910.1728

\bibitem[Geller \& Huchra(1989)]{gel89}
Geller, M. J., \& Huchra, J. P. 1989, Science, 246, 897

\bibitem[Giavalisco et al.(1998)]{gia98}
Giavalisco, M. et al. 1998, ApJ, 503, 543

\bibitem[Gott et al.(2005)]{got05}
Gott, J. R., III, et al. 2005, ApJ, 624, 463

\bibitem[Gunn et al.(1998)]{gun98}
Gunn, J. E., et al. 1998, AJ, 116, 3040

\bibitem[Gunn et al.(2006)]{gun06}
Gunn, J. E., et al. 2006, AJ, 131, 2332
 
\bibitem[Ivezi{\'c} et al.(2004)]{ive04}
Ivezi{\'c}, {\v Z}., et al. 2004, Astron. Nachr., 325, 583

\bibitem[Lacoste et al.(2005)]{lac05} 
Lacoste, C., Descombes, X., \& Zerubia, J. 2005, IEEE Trans. 
Pattern Analysis and Machine Intelligence, 27, 1568

\bibitem[Le F{\`e}vre et al.(2005)]{lef05}
Le F{\`e}vre, O., et al. 2005, A\&A, 439, 877

\bibitem[Meneux et al.(2006)]{men06} 
Meneux, B. et al. 2006, A\&A, 452, 387
 
\bibitem[Moody, Turner \& Gott(1983)]{moo83}
Moody, J. E., Turner, E. L., \& Gott, J. R. 1983, ApJ, 273, 16

\bibitem[Novikov, Colombi \& Dor{\'e}(2006)]{nov06}
Novikov, D., Colombi, S., \& Dor{\'e}, O. 2006, MNRAS, 366, 1201

\bibitem[Ouchi et al.(2004)]{ouc04}
Ouchi, M. et al. 2004, ApJ, 611, 685

\bibitem[Phleps \& Meisenheimer(2003)]{phl03}
Phleps, S. \& Meisenheimer, K. 2003, A\&A, 407, 855

\bibitem[Rawlings et al.(2004)]{raw04}
Rawlings, S. et al. 2004, arXiv:astro-ph/0409479

\bibitem[Schechter(1976)]{sch76}
Schechter, P. 1976, ApJ, 203, 297

\bibitem[Schlegel et al.(2009)]{sch09}
Schlegel, D. J. et al. 2009, arXiv:0904.0468

\bibitem[Smith et al.(2002)]{smi02}
Smith J. A. et al. 2002, AJ, 123, 2121
  
\bibitem[Sousbie et al.(2008a)]{sou08a}
Sousbie, T., Pichon, C., Colombi, S., Novikov, D., \& Pogosyan, D. 2008a, 
MNRAS, 383, 1655

\bibitem[Sousbie et al.(2008b)]{sou08b}
Sousbie, T., Pichon, C., Courtois, H., Colombi, S., \& Novikov, D. 2008b, 
ApJ, 672, L1

\bibitem[Soubie, Colombi \& Pichon(2009)]{sou09}
Sousbie, T., Colombi, S., \& Pichon, C. 2009, MNRAS, 393, 457

\bibitem[Stoica et al.(2005)]{sto05}
Stoica, R. S., Martinez, V. J., Mateu, J., \& Saar, E. 2005, ApJ, 434, 423

\bibitem[Stoica et al.(2007)]{sto07}
Stoica, R. S., Martinez, V. J., \& Saar, E. 2007, Journal of the Royal 
Statistical Society : Series C (Applied Statistics), 55, 189

\bibitem[Stoughton et al.(2002)]{sto02}
Stoughton C., et al. 2002, AJ, 123, 485 

\bibitem[Strauss et al.(2002)]{str02}
Strauss, M A., et al. 2002, AJ, 124, 1810

\bibitem[Willmer et al.(2006)]{wil06}
Willmer, C. N. A. et al. 2006, ApJ, 647, 853

\bibitem[York et al.(2000)]{yor00} 
York D. G. et al. 2000, AJ, 120, 1579

\bibitem[Zel'Dovich(1970)]{zel70}
Zel'Dovich Y. B. 1970, A\&A, 5, 84

\end{thebibliography}
\end{document}